\documentclass[submission]{eptcs}

\usepackage{iftex}
\usepackage{listings,xspace,xcolor}
\usepackage{url}             % 
\usepackage{hyperref}        %

\usepackage{amsfonts}
\usepackage{amsmath}
\usepackage{amssymb}
\usepackage{graphicx}
\graphicspath{ {./images/} }
\newcommand{\plugin}[1]{\textsf{#1}\index{#1 \emph{(Frama-C plug-in)}}\xspace}

\newcommand{\cadna}{\textsf{Cadna}\xspace}
\newcommand{\verrou}{\textsf{Verrou}\xspace}
\newcommand{\verificarlo}{\textsf{Verificarlo}\xspace}
\newcommand{\fldlib}{\textsf{FLDLib}\xspace}

\newcommand{\from}{\plugin{From}}

\lstdefinelanguage{ACSL}{%
	escapechar={},
	literate=,
	breaklines=false,
	morekeywords={
        admit,allocates,assert,assigns,assumes,axiom,axiomatic,behavior,behaviors,
        boolean,breaks,case,check,complete,continues,
        unroll,
        decreases,disjoint,ensures,
	exits,frees,exit_behavior,ghost,global,impact,inductive,integer,invariant,
        lemma,logic,loop,model,relational,pragma,predicate,reads,real,requires,returns,sizeof,
        slice,strong,struct,terminates,type,union,variant,typically,
	Init,Here,LoopCurrent,LoopEntry,Old,Pre,Post,
        \\allocation,\\allocable,\\automatic,
	\\at,\\base_addr,\\block_length,\\dangling,\\dynamic,\\empty,\\exists,\\forall,\\freeable,
        \\fresh,\\from,\\ghost,\\in,\\initialized,\\inter,\\lambda,\\let,
	\\nothing,\\null,\\object_pointer,\\offset,\\old,\\register,\\result,
        \\separated,\\static,\\type,\\typeof,\\union,\\valid,\\valid_function,
	\\valid_read,\\with,
        \\abs,\\acos,\\add\_double\\add\_float,\\asin,\\atan,\\atan2,\\ceil,
        \\concat,\\Cons,\\cos,\\cosh,
        \\div\_double,\\div\_float,\\Double,\\Down,\\e,\\false,\\floor,
        \\eq\_double,\\eq\_float,\\exact,\\exp,\\fabs,\\fmod,\\ge\_double,
        \\ge\_float,\\gt\_double,\\gt\_float,\\hypot,
        \\is\_finite,\\is\_infinite,\\is\_minus\_infinity,\\is\_NaN,
        \\labs,\\le\_double,\\le\_float,\\length,\\list,\\log,\\log10,
        \\lt\_double,\\lt\_float,\\max,\\min,
        \\minus\_infinity,\\mul\_double,\\mul\_float,\\model,\\NaN,\\ne\_double,\\ne\_float,
        \\NearestAway,\\NearestEven,\\neg\_double,\\neg_float,\\Negative,\\nth,
        \\no\_overflow\_single,\\no\_overflow\_double,
        \\Nil,\\numof,\\pointer\_comparable,\\pi,\\plus\_infinity,\\Positive,\\pow,
        \\product,\\repeat,\\round\_double,\\round\_error,
        \\round\_float,\\Quad,\\sign,\\sin,\\sinh,\\Single,\\sqrt,\\sub\_double,
        \\sub\_float,\\sum,\\tan,\\tanh,
        \\total\_error,\\ToZero,\\true,\\truncate,\\Up
        },
	alsoletter={\\-},
	morecomment=[l]{//}
}
\lstMakeShortInline[language=C,alsolanguage=ACSL]"
\lstnewenvironment{codeACSL}{\lstset{language=ACSL,stepnumber=0}}{\smallskip}
\lstdefinestyle{c}{language={[ANSI]C},%
	alsolanguage=ACSL,%
	moredelim={**[s]{/*@}{*/}},%
	moredelim={*[l]{//@}},%
}
\def\+{\discretionary{}{}{}}

\newcommand{\commentGB}[1]{\color{red}((GB: #1))\color{black}}
\newcommand{\commentNK}[1]{\color{violet}((NK: #1))\color{black}}
\newcommand{\commentFV}[1]{\color{blue}((FV: #1))\color{black}}
\renewcommand{\commentGB}[1]{}
\renewcommand{\commentNK}[1]{}
\renewcommand{\commentFV}[1]{}

\ifpdf
  \usepackage{underscore}         % Only needed if you use pdflatex.
  \usepackage[T1]{fontenc}        % Recommended with pdflatex
\else
  \usepackage{breakurl}           % Not needed if you use pdflatex only.
\fi

\title{A Case Study on Numerical Analysis of a Path Computation Algorithm}
\author{Grégoire Boussu
\institute{Thales Research \& Technology\\
Palaiseau, France}
\email{gregoire.boussu@thalesgroup.com}
\and
Nikolai Kosmatov %\orcid{0000-0003-1557-2813}
\institute{Thales Research \& Technology\\
Palaiseau, France}
\email{nikolai.kosmatov@thalesgroup.com}
\and
Franck Védrine
\institute{Université Paris-Saclay, CEA, List \\
Palaiseau, France}
\email{franck.vedrine@cea.fr}
}
\def\titlerunning{Numerical Analysis of a Path Computation Algorithm}
\def\authorrunning{G. Boussu, N. Kosmatov and F. Védrine}
\begin{document}
\maketitle

\begin{abstract}
Lack of numerical precision in control software~--- in particular, 
related to trajectory computation~--- can lead to 
incorrect results with costly or even catastrophic consequences.
Various tools have been proposed to analyze the precision of program computations.
This paper presents a case study on numerical analysis of an industrial
implementation of the fast marching algorithm,  a popular path computation algorithm 
frequently used for trajectory computation. 
We briefly describe the selected tools, present the applied methodology, 
highlight some attention points, summarize the results and outline 
future work directions.
\end{abstract}

\section{Introduction}
\label{sec:intro}
% !TeX root = ./main.tex

Numerical precision of algorithms has become an important concern
for modern critical software.
Accumulation of rounding errors can lead to serious 
issues in programs involving
floating-point numbers. Such accumulated errors can significantly affect the accuracy of computations and lead to
incorrect results.
Even for a mathematically correct algorithm~--- considered in real numbers~--- its computer implementation can give inaccurate or incorrect results if this implementation does not  properly take into consideration numerical precision aspects
of the resulting computation in floating-point numbers.
In critical software, in particular in control software related to trajectory computation, lack of numerical precision can lead to 
incorrect results with costly or even catastrophic consequences. Well-known examples include 
the Patriot missile failure in 1991\footnote{See  \url{https://www-users.cse.umn.edu/~arnold/disasters/Patriot-dharan-skeel-siam.pdf}.} and
the crash of Ariane 5 in 1996\footnote{See  \url{https://www-users.cse.umn.edu/~arnold/disasters/ariane5rep.html}.}.

The fast marching algorithm~\cite{Sethian96} is  a popular path computation algorithm
frequently used for trajectory computation in autonomous systems.
It  answers  the question of which path is optimal between two given nodes, that is, has the shortest time or, more generally, the smallest weight.   
The algorithm works in two steps. 
A first step performs a forward wave front propagation from the given origin point, computing the time the wave front will take to reach each point (of the plan, or grid, or graph). 
A second step uses the resulting computation to perform a backward 
propagation from the final point to the origin point in order to compute an optimal path.
This algorithm has various applications for trajectory computation
and image segmentation. The purpose of this work is to investigate the numerical precision of an industrial implementation by Thales of this
algorithm over a discrete grid. 

Numerical analysis of such trajectory computation algorithms is a very 
challenging and time-consu\-ming task. Execution paths in the code are typically very long and go through many instructions. Each of them can have an impact on precision and robustness of the algorithm.
Indeed, such a path can go through many \emph{unstable branches}, that is, branches after a conditional expression for which 
a small imprecision of computation
or a small variation of input values can change the truth value of the condition and lead to another branch in the code (ex: \textit{else} instead of \textit{then} branch of a conditional statement, or one more loop iteration), possibly impacting the rest of the algorithm. 
Rounding a floating-point number to an integer can have a similar  impact when the resulting integer is later used in the code: if a value around 100.0 can be rounded to 99 or 100, it can
potentially have a significant impact.  
Moreover, since the algorithm simulates a continuous real space by a discrete grid, a deviation at one node can easily involve different
nodes and thus lead to a quite different result.

Various techniques and tools have been proposed to analyze the precision of program computations. They include dynamic analysis and static analysis techniques. 
In this  work, we use three popular numerical analysis tools:
\cadna~\cite{cadna}  and \verrou~\cite{verrou} 
realizing (possibly unsound) dynamic analysis, and
\fldlib~\cite{fldlib}  performing 
%sound approximation of the imprecision using
a combination of sound abstract interpretation  
and dynamic path exploration.

\paragraph{Contributions.}
This paper presents a case study on numerical analysis of an industrial
implementation of the fast marching algorithm.
While the considered implementation is currently not publicly available, the underlying algorithm is classic,
therefore we believe that 
%we believe that 
the presented methodology and findings can be of interest
for other implementations of similar (and possibly other) 
%path computation 
algorithms.
We briefly describe the selected tools, present the applied methodology combining several tools, 
highlight some attention points, summarize the results and outline
ongoing and future work directions.

\paragraph{Outline.} Section~\ref{sec:algorithm} presents the considered algorithm. Section~\ref{sec:verification} describes
the verification methodology, the selected tools and our findings. 
Section~\ref{sec:conclusion} provides a conclusion and 
future work perspectives.
%\cite{titoloMFM20}
%\cite{kirchner15fac}

\section{The Verification Target: the Fast Marching Algorithm}
\label{sec:algorithm}
% !TeX root = ./main.tex

%\commentNK{Algorithm presentation here} 

This section provides a simplified presentation of the problem and 
the implemented algorithm without giving all technical and theoretical details (which are not mandatory for understanding the paper). For a more thorough description of theory behind the Fast Marching Algorithm, one may refer to \cite{SETHIAN2001503}.

The problem under consideration for the study is named the \emph{minimum-cost path problem}. 
On a finite graph with weighted edges, this problem can be stated as follows: which path to take between two specified vertices so that the sum of weights along this path is the lowest among all possible paths between the two nodes. 
When the weight is (seen as) the distance between the nodes, 
this problem is also called the \emph{shortest path problem}, and the cost is (seen as) the time to reach the point. 
Different algorithms exist to solve the shortest path problem ({e.g.} Dijkstra, Bellman-Ford). 

In our case, we are interested in the definition of the \emph{minimum-cost path problem} in the continuous case: let us consider the problem in $\mathbb{R}^n$. A cost density function $\tau \, \colon \, \mathbb{R}^n \to (0, \infty)$ gives the cost at each point of the space. The \emph{minimum-cost path problem} between $A$ and $B$, two points in $\mathbb{R}^n$, is to find a path $c(s): [0, \infty) \to \mathbb{R}^n$ that minimizes the cumulative cost (often interpreted as the arrival time) from $A$ to $B$. The \emph{cumulative cost} for a path $c$ between $A$ and $M$ is:
$$ T_c(M) = \int_{0}^{l} \tau(c(s)) \, ds $$
where $l$ is the length of path $c$ between $A$ and $M$, $c(0) = A$ and $c(l) = M$. Therefore, $c_{sol}$ is a solution to the problem if and only if $c_{sol} \in \mathcal{C}$ where $\mathcal{C}$ is the set of all paths $c$ between $A$ and $B$ with the minimum $T_c(B)$. 

As stated in \cite{Sethian96}, if $c_{sol}$ is a solution to the problem, it satisfies the equation below, named the Eikonal equation, for all $M \in c_{sol}$:
$$|| \nabla T_{c_{sol}}(M) || = \tau(M)$$
where $\nabla$ denotes the gradient, and $ || \cdot ||$ denotes the Euclidean norm. 
This equation is, in particular, a way to describe the propagation of a wave front initiated in point $A$. The front speed at point $M$ is given by $1/\tau(M)$, and $T_{c_{sol}}(M)$ is the time of arrival of the front from point $A$ to point $M$.

In the presented definition of the problem, the value of $\tau$ at a given point depends only on the point's location. This case is qualified as \emph{isotropic}. If $\tau$ also depends on the direction of the path at the point, the cost function is \emph{anisotropic}.
A method to solve this equation in the case of an isotropic problem discretized on a Cartesian grid was proposed by Sethian in 1995 \cite{Sethian96} and has become the starting point for many extensions. This method, named \emph{fast marching method} (FMM), shares many aspects with Dijsktra's algorithm. Once the equation is solved for all points of the grid, a second step is necessary to figure out the (or one of the) optimal path solution(s) to the {minimum-cost path problem}. We will go over the two steps sequentially.

\subsection{First Step: Solving the Eikonal Equation}
\label{sec:eikonal-algo}
\begin{figure}[t]
\includegraphics[scale=0.7]{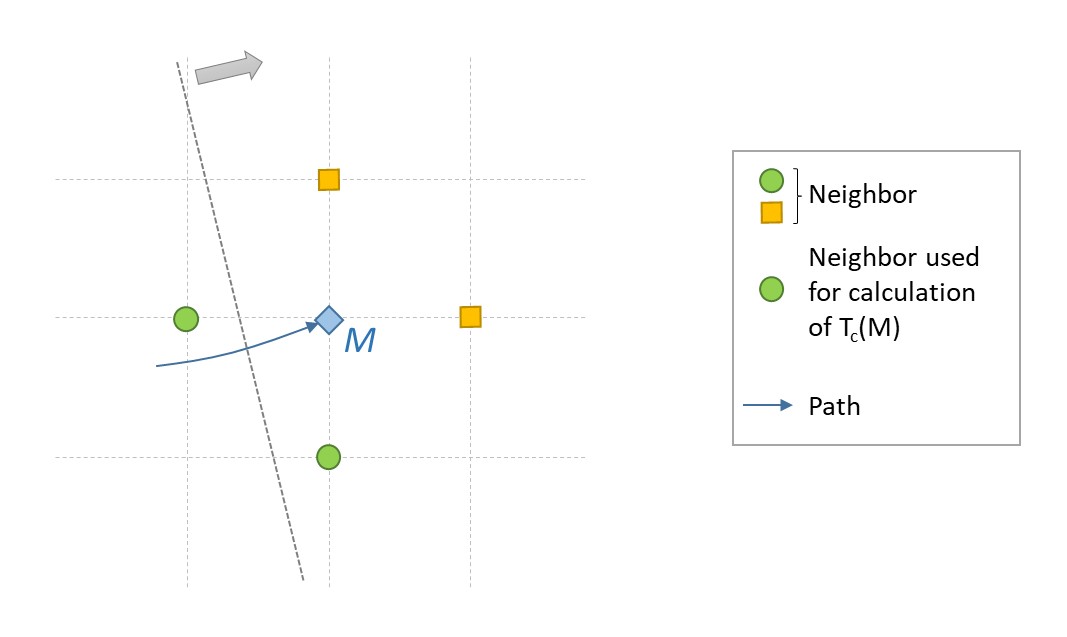}
\centering
\caption{Neighbors selected for calculation of $T_c(M)$}
\label{fig:selected}
\end{figure}

Basically, given a grid and a starting point $A$ of the grid, Sethian's method allows one to calculate the arrival time $T_c(M)$ to any point $M$ of the grid over a minimum-cost path $c$ starting from point $A$. Each point $M$ of the grid has 4 neighbors as shown in Fig.~\ref{fig:selected}. Based on a relevant approximation scheme, $T_c(M)$ is calculated considering the possibilities that the wave about to reach the point M comes from North-East (with a contribution from the neighbors above and on the right), or South-East (with a contribution from the neighbors below and on the right), or South-West (with a contribution from the neighbors below and on the left, as shown in Fig.~\ref{fig:selected}) or North-West (with a contribution from the neighbors above and on the left). Starting the algorithm with $T_c(A) = 0$ and $T_c(M) = \infty \,$ for all $\, M \neq A$, and picking the next point M to study in an appropriate order, the process progressively computes the arrival time $T_c(M)$ (or more precisely, its approximation due to the discretization) for all points $M$ of the grid.

\begin{figure}[t]
\includegraphics[scale=0.7]{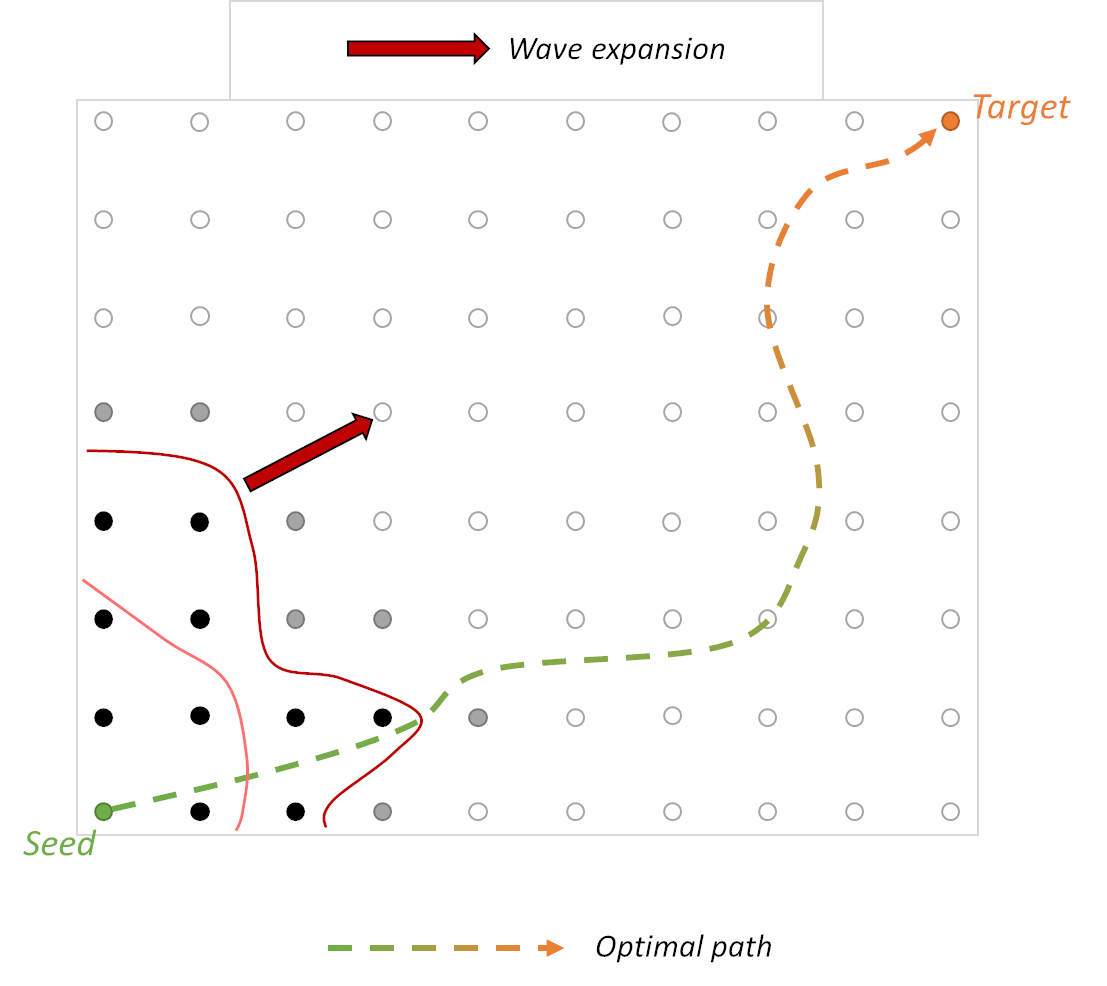}
\centering
\caption{Propagation of the wave front}
\label{fig:propagation}
\end{figure}

We can further explain the process
using the interpretation of the equation with a wave front, illustrated in Fig.~\ref{fig:propagation}. The black points of the grid have their final value $T_c(M)$ computed, the gray ones have a tentative value $T_c(M)$ computed, and the white ones still have $T_c(M) = \infty$, as set at the initialization step.

The set of gray points is named the \emph{narrow band}. Intuitively, %by a given moment of time, 
at each step of the algorithm,
the black points have already been reached by the wave, and
at least one of the gray points will be reached next, before any of the white points will be reached. Just like in a classic implementation of Dijkstra's algorithm, a priority queue is used to store the gray points. When a point $M$ is selected from the queue, its neighbors $M'$ enter the queue (if they were not already part of it) and get their arrival time values $T_c(M')$ calculated or updated based on  $T_c(M)$. The next point $M$ to be selected in the queue is the one with the lowest tentative arrival time $T_c(M)$. When selected, such a gray point gets its tentative value $T_c(M)$ turned to the final one, and the point itself is removed from the queue and labeled as black. 
Intuitively, since the tentative arrival time of the wave to this point is the smallest one among the gray nodes, it cannot be reached even faster through some other node (for which the arrival time will necessarily be bigger) hence the computed arrival time to it is final.
At the beginning of the algorithm, the priority queue is initialized with $A$.

The fast marching method has two interesting features:
\begin{itemize}
    \item It is efficient in terms of computational complexity. Indeed, its complexity is similar to that of Dijsktra's algorithm, and is of $\mathcal{O}(n \lg{(n)})$, with $n$ being the number of points of the grid.
    \item It can be proven that the FMM produces a solution that satisfies everywhere the discrete version of the Eikonal equation, leading to an approximation of its so-called \emph{viscosity} solution (see for example \cite{crandall1992usersguideviscositysolutions} on viscosity solutions). So when the grid spacing tends to 0, the solution provided by the FMM algorithm tends to the continuous solution of the equation.
\end{itemize}

Though, as $T_c(M)$ is calculated based on the $T_c$ of the neighbors of $M$, the calculation errors may propagate over the entire grid. Added up, these errors may lead to a discrepancy for points far from point $A$ and impact the precision of the global result expected from using FMM. Studying the order of magnitude of this discrepancy is of great interest to be confident in the implementation of the algorithm.

\subsection{Second Step: Finding an Optimal Path by a Backward Propagation}
\label{sec:backpath-algo}
The theory provides a way to find an optimal path, thanks to a property of such a path: its direction is always normal to the wave front~\cite{bellman1965dynamic}. To produce the result, a so-called back-propagation from the final point (supposed to be on the grid) is realized, based on a gradient descent following the direction perpendicular to the wave front curve. 

Though, once the  arrival time values $T_c(M)$ are calculated for each point of the grid, we are still in a discrete space and the gradient calculation is not straightforward. The gradient descent can be approximated by selecting for each point its predecessor among the neighbors, the appropriate one being the (or one of the) neighbor(s) with the lowest $T_c(M)$. But this approach leads to a path made of following segments that can be perpendicular one to the next. Moreover,  aggregating the length of each segment of the path will generally lead to a value overestimated compared to the optimal path length in a continuous space. It would be preferable to provide a visually smooth path with its length approximating the length of the viscosity solution of the Eikonal equation.

\begin{figure}[t]
\includegraphics[scale=0.7]{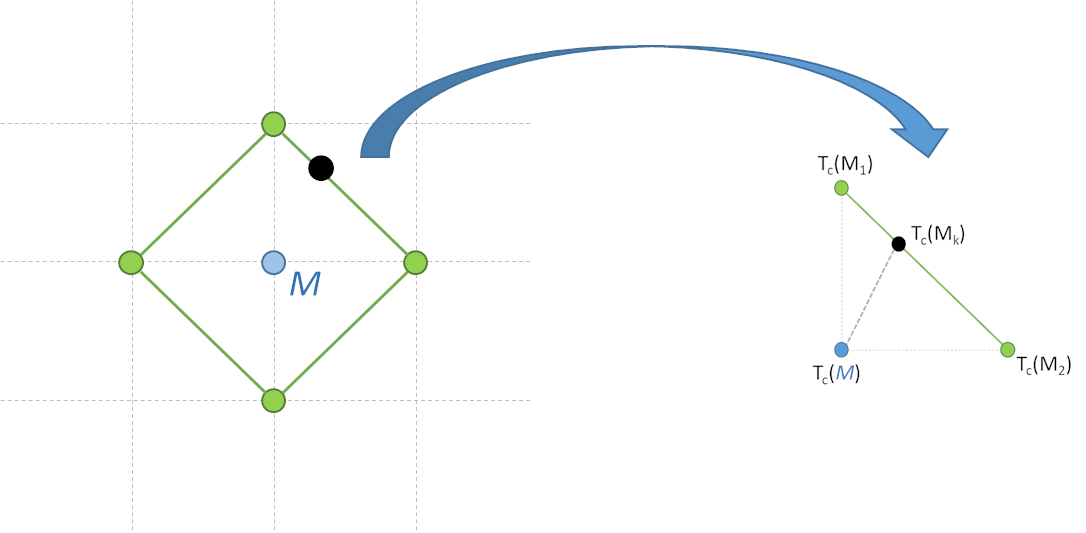}
\centering
\caption{Calculation of pseudo-gradient on a segment}
\label{fig:pseudogradient}
\end{figure}

Such an alternative can be implemented with the following approach: starting from the final point of the path, a \textit{pseudo-}gradient is calculated on each segment around this point, as shown in Fig.~\ref{fig:pseudogradient}. The best point on the segments, {i.e.} the point (or one of the points) minimizing $\Delta T_c / distance$ (that is, maximizing the speed of the wave) is selected as the previous point of the approximated path. A similar approach is taken to find out the best point when the gradient is calculated from any point in the middle of a segment. This leads to a much smoother path, whose length provides a good approximation of the expected length of the viscosity solution. 

Just as in the case of the fast marching algorithm, the calculation is made one point after the other. Therefore, the calculation errors due to the implementation can lead to an aggregated discrepancy. An analysis of sensitivity of the implementation to these errors is thus required.

\subsection{Applications of These Algorithms}
Many fields of application exist for these algorithms. The first is of course related to path calculation leading to the shortest time between two points, considering the speed of the mobile agent depending on its position in an area. A less obvious application could be image segmentation~\cite{ImageSegmentation}. To allow for different and more specific situations, many extensions to the method have been developed. To name a few: the possibility to deal with anisotropic costs~\cite{KimmelComputing1998}, or with time-dependent costs with no restriction on sign~\cite{carlini2008convergence}, or the extension taking into consideration constraints like minimum turning radius of the moving agent~\cite{mirebeau2023massively}.

\begin{figure}[t]
\includegraphics[scale=0.35]{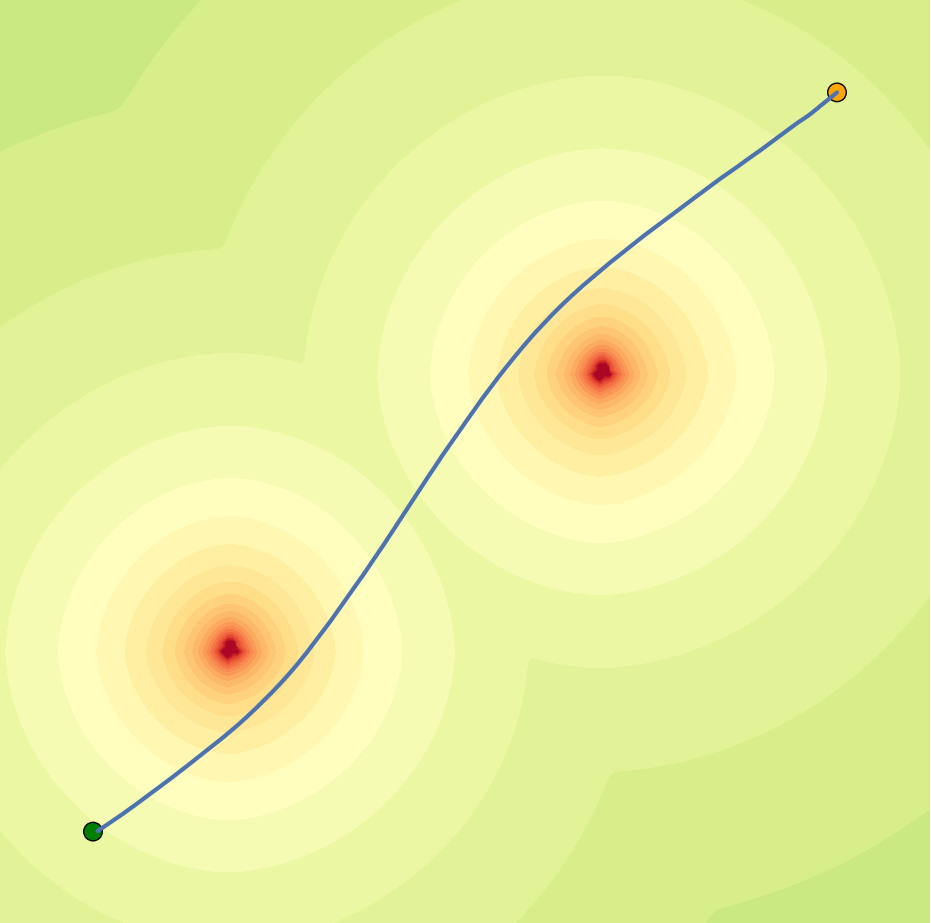}
\centering
\caption{Result of calculation of a path avoiding turbulence areas}
\label{fig:path}
\end{figure}

In our case, the fast marching method is a general approach to produce paths optimizing any kinds of criteria (or a mix of criteria). The most straightforward situation would be to aim at minimizing time to reach a location, while the area through which we can move is made of danger-free zones where the speed can be high, and others surrounded by dangers (mountains, ...) where the velocity should be reduced. Let us consider another example where time is not the criterion to optimize: the pilot of a plane  wishes to avoid turbulence areas ahead (considered as static). He may want to find a good balance between disturbance due to very strong winds and the additional distance incurred by avoiding these areas. If the plane has a steady cruising speed, by defining the cost function $\tau$ with high values in the center of the turbulence areas and decreasing values towards the outside, the fast marching method can provide an appropriate path to follow (see Fig.~\ref{fig:path}).

For use cases where a lack of precision can generate additional risks (e.g. air traffic, autonomous drones), aggregated calculation errors can significantly impact the result of the computation. This concern motivated the current study.

\section{The Verification Approach and Results}
\label{sec:verification}
% !TeX root = ./main.tex

%\commentNK{Verification approach and results here. Maybe split into several sections?}

\mbox{}

The target implementation of the path computation algorithm contains more than 6,200 lines of C++ code and provides several test cases.
They include realistic test cases over a square grid with 200x200 nodes and obstacles simulated by higher weights, as illustrated by Fig.~\ref{fig:path}. 
Internally, the code uses some C++ STL  (Standard Template Library) containers, like vectors, maps and priority queues.
So, formal numerical analysis of this implementation and its adequacy with respect to the underlying
mathematical formulas within a short period of time requires concentrating on successive research questions:

\begin{itemize}
\item[RQ1:] What is the accuracy of the computed path cost? 
\item[RQ2:] Is the computation robust (meaning that  a small perturbation of inputs leads only to a small variation of outputs)?
\item[RQ3:] How does the computed path compare to the path obtained on a more or less precise grid (say, with twice more or twice fewer points on each side)?
%\item[RQ3:] how does the computed path compare to the result of another path optimization algorithm?
%\item[RQ4:] how does the computed path compare to the mathematical formula that underlies the implementation?
\end{itemize}

%\commentFV{We could suppress some questions if we have no result for them? Or the conclusion
%  can bring some answers for these questions?}
%\commentNK{I think, we should better keep the questions we address. The others can be mentioned in the future work at the end?}

This paper focuses on RQ1 and RQ2, while RQ3 is left for future work.  
To address these questions, we decided to apply the following methodology that was successfully
applied earlier on some simpler numerical use-cases, except the last item that is more related to
deductive verification:
\begin{itemize}
\item instrument the tests with different numerical analysis libraries to identify the difficulties
  in obtaining relevant analysis results and then refine the verification objectives,
  such as accuracy requirements;
\item enlarge the tests into analysis scenarios to check whether the analysis scales up and still provides
  precise results. Fine-grained analysis scenarios typically replace concrete input values by very small
  input intervals and then apply conservative interval operations;  larger analysis scenarios
  can also be considered;
\item apply modular formal verification to the components of the  target implementation and
  assemble the reasoning results to provide a proof of global correctness.
\end{itemize}

After presenting the common instrumentation in the next section, we will apply \cadna to address
RQ1, \verrou to address RQ1 and RQ2 by comparing with \cadna results, and \fldlib to address
RQ1 and RQ2 to investigate the unstable branches that may have a significant impact on the robustness results.

\subsection{A Common Instrumentation Mechanism for Different Verifications}
\label{sec:instrumentation}

The mechanism for building the target implementation of the fast marching algorithm uses the
\texttt{cmake} tool; so a slight
modification of the file \texttt{CMakeLists.txt} enables adding some new executable targets
compiled with specific compilation flags.
This feature enables the source code to be easily compiled with analysis libraries into a single executable.
For the verification purposes, this instrumentation mechanism competes with
abstract interpreters when the abstractions to be used are generic (intervals, affine forms).
But our case study also requires the
elaboration of specific abstractions. So, to quickly explore and debug these newly created abstractions,
an instrumentation based on C++ operator overloading
and template/macros mechanisms seemed to us more efficient than using an existing generic
abstract interpreter.

Our default instrumentation mechanism replaces \texttt{double} and \texttt{float} types with
data structures
carrying analysis information like accuracy, as it is often done by instrumentation libraries~\cite{cadna,
fldlib}. Such data structures
implement an overload for the arithmetic  operations
(\texttt{+}, \texttt{-}, \texttt{*}, \texttt{/}, \texttt{pow})
to infer numerical properties like the accumulation of round-off errors in numerical computations.
Integer types like \texttt{int} or \texttt{unsigned int} are not instrumented by default.
But, the source code can use explicit intrumentation for these types by replacing \texttt{int} by
\texttt{EnhancedInteger<int>} whenever it makes sense for some \texttt{EnhancedInteger} template
class to define. The C++ compiler helps then to statically
propagate these custom types on the source code since an operation manipulating
\texttt{int} and \texttt{EnhancedInteger<int>} generates an error if its result is
assigned to an \texttt{int} and not to an \texttt{EnhancedInteger<int>}.

For each analysis target, the file \texttt{CMakeLists.txt} adds specific compilation flags
like \linebreak \texttt{-I.../analysis_include -include std_header.h -DFLOAT_MY_ANALYSIS} to build
the target. \linebreak The directory \texttt{.../analysis_include} contains the file \texttt{std_header.h}
that conditionally loads the appropriate analysis data structures for the flag
\texttt{FLOAT_MY_ANALYSIS} and replaces the \texttt{double} type with the macro \texttt{double} 
defined by \texttt{\#define double EnhancedFloatingPoint<double>}.

\mbox{}

The research questions stated in the beginning of this section systematically compare two or more executions.
These executions can (and do) follow different control flows (that is, different execution paths) in the target program.
%\commentNK{Not clear what synchronous/asynchronous means in the next sentense. Explain? E.g.:
% Moreover, applied analysis can be synchronous or asynchronous, depending if it is executed
% simultaneously with the execution of the instrumented program or not.?}\commentFV{our analyses
%are always executed with the perturbed execution of the instrumented program: the execution is
%the master and the analysis semantics is a slave of the execution}
In our experiments, we instrument the code with three different strategies:
\begin{enumerate}
\item A single run of a synchronous analysis with a single control flow: this single analysis run
  propagates complete analysis information for every variable at every point of the execution
  path until the end of the program.  
\item Multiple runs of asynchronous analyses with a single control flow: a run propagates
  partial analysis information until the end of the program. With multiple runs, the user can
  compute the analysis result as a model from the correlated input/output data.
\item A single run of a synchronous analysis with multiple control flows: this single analysis
  run propagates complete analysis information and thanks to additional local loops, it covers
  all possible execution paths (which corrects a weakness of Strategy 1 with an additional instrumentation
  and execution cost).
\end{enumerate}
For the last strategy, 
%the \texttt{std_header.h} file provides 
we use 
\texttt{SPLIT}/\texttt{MERGE} macros introduced 
and used by \fldlib~\cite{fldlib}. 
A pair of such macros (\texttt{SPLIT} and 
\texttt{MERGE}) define a so-called \texttt{SPLIT}/\texttt{MERGE}
section: it expands into a local loop that iterates over all the reachable control flows of the \texttt{SPLIT}/\texttt{MERGE}
section in order to analyze them one after another. 
A local memory defined in
the \texttt{SPLIT} macro saves the memory before the section and restores it at the beginning
of the loop body for an exploration of a new control flow. 
%The \texttt{MERGE} macro then synchronizes the analysis results at the end of this loop.
\texttt{SPLIT} also saves a control flow
identifier for an exploration of a new execution path of the section. It then increments this identifier to cover another execution path in the next loop
iteration. 
At the end of the loop, \texttt{MERGE} incrementally synchronizes the results of
the local analyses to create a single analysis summary per variable. The analysis is then continued with this
summary until the end of the program.

\medskip

Beyond these generic principles, the instrumentation may encounter some problems listed below,
which may require minor adaptations to the source code for analysis purposes. In
practice, the first two problems are absent in the modern C++ implementation of the fast marching
algorithm.

\begin{itemize}
\item dynamic allocations with C functions \texttt{malloc} and \texttt{free} should be replaced
  by C++ \texttt{new} operator with smart pointers (or \texttt{delete}):
  \texttt{EnhancedFloatingPoint} often has non-trivial
  constructor and destructor and the \texttt{malloc} and \texttt{free}
  functions do not call them unlike \texttt{new} and \texttt{delete}.
\item C functions with variable number of arguments and specialized format specifiers
  (such as \texttt{scanf} and \texttt{printf(``\%e'', \ldots)})
  should be replaced with \texttt{std::cout}, \texttt{std::cin} calls because \texttt{``\%e''}
  does not recognise \texttt{EnhancedFloatingPoint}.
\item In a divide-and-conquer analysis approach, we typically instrument certain parts of the code and
  leave others unchanged. But, replacing \texttt{int} with
  \texttt{EnhancedInteger<int>} also generates many other replacements. In the case of the
  fast marching algorithm, the forward propagation part (Sect.~\ref{sec:eikonal-algo})
  and the backward propagation for path generation (Sect.~\ref{sec:backpath-algo}) share some common methods.
  However, replacing \texttt{int} with \texttt{EnhancedInteger<int>} is only required in the backward
  path generation. In this case we rename the original method as a template method in the private
  section of the class. Then, we duplicate the public method, one with \texttt{int} arguments
  and the other with \texttt{EnhancedInteger<int>} arguments. The bodies of the original method
  and its duplication just call
  the template private method. From the caller's perspective, the C++ ``name lookup''
  generally generates correct calls.
\item The second argument of binary operators whose first argument is 
  of type
  \texttt{EnhancedInteger<int>} may be \texttt{int}, \texttt{unsigned}, \texttt{double},
  \texttt{EnhancedFloatingPoint<double>}. The instrumentation needs precise overloaded
  operators to be called by all the constructs of the source code. In C++-03, providing
  an interface for this instrumentation that correctly connects the source operator with
  the correct overloading for all type combinations 
  was a very complex task and ultimately produced a resulting interface that was difficult to manage. That is
  probably the reason why \cadna~2.1 does not support \texttt{long double}. Then, the
  \textit{SFINAE} (Substitution failure is not an error)~\cite{wik:sfinae} feature allows
  the definition of such a robust interface, but this remains very technical with
  maintainance difficulties.
  Our libraries use recent C++-20 concepts to manage this \texttt{class} interactions, which
  makes the instrumentation more robust.
%  \commentNK{This last item is not clear to me.}
%  \commentNK{I hope the new text is clearer.}
\end{itemize}

\subsection{Results of the Approach based on Dynamic Analyses}

To address RQ1, the objectives of the first analyses are
\begin{itemize}
\item ensuring that the code can be instrumented with dynamic analysis libraries (that are generally simple to use from the
  instrumentation point of view),
\item obtaining initial quantitative accuracy properties to be refined later with more complex
  analyses,
\item evaluating the robustness of the implementation: a small perturbation of input
  data should generate a small deviation in the outputs. If the implementation is not
  robust, we have no chance of proving formal functional properties, such as ``the results depend in a limited way on the size of the grid''.
\end{itemize}

Dynamic analysis
with stochastic arithmetic meets these goals; that is why we
use it as a first approach. To do this, we couple the instrumentation mechanism described in the previous section
with stochastic analysis libraries in order to obtain accuracy and robustness results
without any modification to the source code. Such analyses only require a test case and explore
the impact of minor perturbations on the results after execution of the test scenario.

\begin{figure}
  \centering
  \includegraphics[width=9.8cm]{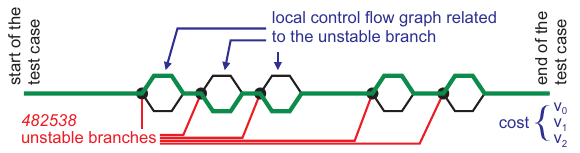}
  \caption{(Simplified) trace of the execution with \cadna (in green), where the cost of the path is evaluated by three values $v_1,$ $v_2,$ $v_3$.}
  \label{fig:cadna-execution}
\end{figure}

The \cadna\footnote{See \url{https://www-pequan.lip6.fr/cadna}}~\cite{cadna} library evaluates the accuracy of a code by propagating three executions in
a single run (synchronous analysis with single control flow). 
This leads to maintaining three values ($v_0$, $v_1$, $v_2$ in Fig.~\ref{fig:cadna-execution})
for each computed variable \textit{var} instead of one. 
To evaluate potential impact of rounding errors, with this library
each floating-point computation involving \textit{var} is dispatched over $v_0$, $v_1$ and $v_2$.
The ideal results are randomly rounded up or down~--- with a
%as $v'_0$, $v'_1$ and $v'_2$ -- three draws 
probability of 50\% for up or 50\% for down~--- instead of using
the deterministic IEEE-754 rules implemented in the processor.
The average of the three values 
provides the expectation of the computed result, while the standard deviation provides an estimate of the accumulation
of round-off errors~\cite{VIGNES1993233}. Such analysis is synchronous: the three executions are \emph{forced} to follow the
same control flow (shown in green in Fig.~\ref{fig:cadna-execution}) and do not evaluate \emph{unstable branches}, for which a possible imprecision can impact the result of  a conditional test (and therefore the branch taken after it). Let us consider a comparison, say $d<d'$ between two \texttt{double} values $d$ and $d'$
instrumented by \cadna. It compares each of the three values
$v_0$, $v_1$, $v_2$ 
obtained for the first value $d$
with the corresponding value of the three values
$v'_0$, $v'_1$, $v'_2$ 
obtained for the second value $d'$. Suppose that the first two comparisons return {true} and the last returns {false}.
\cadna just reports an ``\texttt{UNSTABLE BRANCHING}'' and propagates the last execution
into the {then} branch as well, even if it would naturally execute the {else} branch.

The instrumentation quickly succeeds on the target code with \cadna (thus fulfilling the first objective). The
algorithm computes (in 0.556s) a path of 322 points as well as the cost of the path, stored in variable \texttt{cost}.
Since \texttt{cost} is the value that the algorithm attempts to optimize, we expect it to be robust.
The cost has an average of \texttt{0.392} and a relative error of $1.323\times 10^{-15}$ due to the
accumulation of round-off errors. \cadna also reports the following warnings:

{\scriptsize
\begin{verbatim}
CRITICAL WARNING: the self-validation detects major problem(s). The results are NOT guaranteed.
There are 977194 numerical instabilities
1687 UNSTABLE MULTIPLICATION(S), ...
482538 UNSTABLE BRANCHING(S), 260343 LOSS(ES) OF ACCURACY DUE TO CANCELLATION(S)
\end{verbatim}
}

\noindent
The execution of the test case takes 0.182s and generates a shorter path of 320 points with a cost
of $0.393257$, that is outside the error range computed by \cadna. That confirms~--- as suggested by the warnings~--- that the \cadna results
are not conclusive: unstable branches (not evaluated by \cadna) probably have a major impact on
the path computation and therefore on the robustness of the algorithm.

\mbox{}

\begin{figure}
  \centering
  \includegraphics[width=10.2cm]{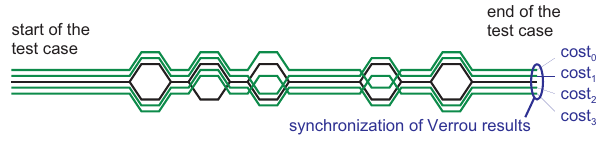}
  \caption{Four (simplified) traces (in green) from the 10 executions with \verrou. Each trace leads to computing a (possibly different) path and its cost}
  \label{fig:verrou-executions}
\end{figure}

The second analysis uses the \verrou\footnote{See \url{https://github.com/edf-hpc/verrou}}~\cite{verrou} tool\footnote{The \verificarlo~\cite{verificarlo} tool (see \url{https://github.com/verificarlo/verificarlo})  
can be expected to produce similar results, but it was not used in this study.}.
\verrou evaluates the accuracy of a code 
during multiple runs
by randomly rounding up or down every
floating-point computation (asynchronous analysis with single control flow). Unlike \cadna,
\verrou does not need additional memory: since the
execution of the program perturbed by \verrou is non-deterministic, multiple runs provide multiple
output values (see Fig.~\ref{fig:verrou-executions}, where we show 
only four traces for readability). The average and the standard deviation of the output
respectively provide the expected stochastic result and an error that is representative
of the accumulation of round-off errors. With ten runs (performed in 16.123s), \verrou provides different
lengths for the optimized path: 324, 307, 308, 307, 315, 315, 314, 317, 312, 300. The average of the ten values of \texttt{cost}
%(see $\textit{cost}_0$, \ldots in Fig.~\ref{fig:verrou-executions})
is evaluated to $0.3922$ and its standard deviation to 
$2.68\times 10^{-4}$.
%\texttt{2.68e-4}.

The error produced by \verrou seems to be more consistent than that of \cadna with respect to the
original IEEE-754 floating-point execution. The multiple runs nevertheless
do not contain the value of cost produced by the test case execution since $0.393257
\gg 0.3922 + 0.000268$. We relaunch the \verrou analysis several times and we systematically obtain
an average and a standard deviation close to these values. That means that the floating-point execution
takes a control path that is distinct from other control paths in terms of their impact on the
\texttt{cost} value. At this point, the use of formal methods appears relevant to further investigate the relative
instability of the floating-point execution.

\subsection{Evaluating the Impact of Perturbations on the Control Flow and the Resulting Path}

To further address RQ1 and investigate RQ2, the third analysis relies on the \fldlib\footnote{See \url{https://github.com/fvedrine/fldlib}} library~\cite{fldlib} to provide 
a sound over-approximation
of the accumulation of round-off errors by maintaining the ideal (in practice, a very precise machine) computation and the floating-point computation in parallel. 
\fldlib relies on SPLIT/MERGE sections (presented above) to analyze
unstable branches by exploring each of the different control flows using abstract interpretation 
and by consolidating the observed results at the end. 
This analysis propagates affine forms for
rounding errors and for the possible values on the test case. The mathematical representation of an
affine form is $\alpha_0 + \sum_{i=1}^n \alpha_i\times \epsilon_i$, where $\alpha_i$ 
are constant coefficients in $\mathbb{R}$ (approximated by floating-point values with a large mantissa)
and $\epsilon_i$ are free variables in the interval $[-1.0, +1.0]$. The error symbols $\epsilon_i$
represent unknown values due to basic approximations of complex computations. The program variables
can share some $\epsilon_i$, which creates linear relationships between some of these variables.
\fldlib also offers advanced features to reduce the
size of the re-executed code with local synchronization annotations (see the \fldlib library documentation and ~\cite{fldlib} for more detail).
%\commentNK{Not sure that paragr is useful, as we have already explained SPLIT/MERGE sections above? I think, that that earlier explanation was clearer. This one is too simplified; it works only if there is just one branch in the section? Also, I cannot follow the text wrt. Fig.7 (where we take the else branch first??? and have several branches???).}\commentFV{Ok to remove it. In Fig 7, the floating-point semantics has only 1 value that takes the else branch. The ideal semantics has a small intervall that can take both branches. The first execution follows the floating-point semantics. The second execution follows the ideal semantics and it starts with the then branch. Since it is the first unstable branch, the $[2^{351296-1}+1, 2^{351296}]$ executions would follow the else branch after having covered all the next branches.}

\begin{figure}
  \centering
  \includegraphics[width=10.6cm]{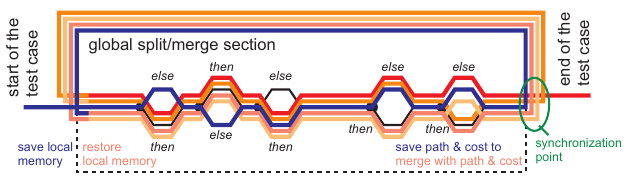}
  \caption{(Simplified) trace of the \fldlib analysis, in which the code of a SPLIT/MERGE section is executed several times for all executable control flows inside it and the results are consolidated at the end}
  \label{fig:fldlib-analysis}
\end{figure}

In practice, we activate the \fldlib analysis after initialization of the mesh.
Therefore, the grid is composed of points with floating-point coordinates considered exact, i.e. without any numerical error. With this assumption,
the analysis only propagates affine forms over 4 execution paths. 
A simplified version of an execution trace of
\fldlib is illustrated by Fig.~\ref{fig:fldlib-analysis}.
After several attempts, we have found the right 
settings: 319 bits for the internal mantissa of the coefficients
of the affine forms and a limit of 30 shared symbols per expression. With an internal mantissa
of 255 bits, the intervals representing the ideal computations are too wide
at the end of the forward wave propagation (see Sect.~\ref{sec:eikonal-algo}). Therefore, the cost associated
with the resulting path is too strongly over-approximated with the interval $[-\infty, +\infty]$: the algorithm performs 
at some moment a division by the distance between two points, and if the
localization of these points is imprecise, a potential division by zero due to interval arithmetic
gives this result. This first \fldlib experimentation (internal mantissa
of 255 bits) is nevertheless interesting because, like
with \cadna and \verrou before,
it also produces a resulting path 
(with $278$
points)
that is different from the path 
produced by the floating-point execution of the test case
(with $320$ points).

\fldlib quickly identifies the location in the source code of a first unstable branching,
for which it explores both branches in floating-point and ideal computation.
%
% {\scriptsize \begin{verbatim}
% main at t-FM2.cpp:168 auto distances_step1 = solver_step1.set_seed(...);
% eks_sethian96::set_seed at eks_sethian96.cpp:27 LOCID seed_id = Mesh.NearestIndex(seed_loc);
% mlb_rectangular::NearestIndex at mlb_rectangular.cpp:385 Nearest_vector(loc.at(0), loc.at(1), i, j);
% mlb_rectangular::Nearest_vector at mlb_rectangular.cpp:92 auto jm = (unsigned int) trunc((y - yMin) / dy);
% \end{verbatim}}
% 
% 
% \commentNK{Not sure the tool output is useful since we do not explain it fully; names of files are not useful to give since the reader does not have access to the code; a higher-level presentation of findings, without citing files and lines, would be preferable}
%
% It is the $12^\textit{\scriptsize th}$ iteration of the loop \texttt{eks_sethian96.cpp:27} for \texttt{loc = (0.245, 0.5)}. 
It concerns the computation of the second instruction of Fig.~\ref{fig:unstable-test}
with the values \texttt{y=0.5}, \texttt{yMin=0}, \texttt{dy=0.005}.

\begin{figure}[tb]
% \centering
\lstset{language=C, keywordstyle=\bfseries\color{black},commentstyle=\color{gray}, showlines=true, moredelim=**[is][\color{gray}]{@}{@}, mathescape, basicstyle=\footnotesize}
\begin{lstlisting}
double jm_res = (y - yMin) / dy;
unsigned int jm = (unsigned int) jm_res;
\end{lstlisting}
\caption{First unstable branch identified with \fldlib}
\label{fig:unstable-test}
\end{figure}

In floating-point semantics, the value of \texttt{jm\_res} is $100$. In the ideal semantics,
the value of \texttt{dy} is $5\times 10^{-3} + 1.0408\times 10^{-18}$ 
since all the constants take the same floating-point value for both semantics; therefore,
the analysis shows that the ideal value of \texttt{jm\_res} belongs to the small interval $[100-2.082\times 10^{-16},
100-2.081\times 10^{-16}]$.
The value of \texttt{jm} is then $100$ in floating-point semantics and $99$ in ideal
semantics, which creates an unstable branching. The analysis then separates the joined control flow
of both semantics into two control flows and explores them separately. These control flows merge at the end of the source code
after the computation of the path and its cost. The merge operation computes the numerical error
from the substraction between the floating-point value and the ideal small interval
of the cost, each value being inferred by the corresponding control flow. The result of this
second \fldlib experimentation (internal mantissa of 319 bits) %is substraction
shows an interesting finding: the unstable branch
has no impact on the cost and on the points of the path, even if the sorted priority queue (see the
gray points of Fig.~\ref{fig:propagation}) is organized differently in the two control flows.
Therefore, the resulting paths both have $320$ points, like the floating-point execution
and they return a relative error of $7.058\times 10^{-16}$ for the cost. %Recall that the error reported by \cadna was $1.323\times 10^{-15}$ with a path different from the execution.
%\commentNK{Conclude MORE EXPLICITLY how this new ERROR of \fldlib compares to the \cadna one?}

The duration of this second \fldlib analysis is 114\,min after a limited exploration of only
$4$ execution flow paths. For one control
flow path, the analysis encounters $351\,296$ unstable branches. Therefore, the estimated
time for the complete analysis would be $2^{351\,296} \times 114\,\mbox{min} / 4$. 
%\commentNK{Not clear why $/ 4$.}
Nevertheless, these preliminary results allow us to
identify the first unstable branches and to show their absence of impact on the computed path and its
cost. 
%\commentNK{Not sure we said what the path variable means. Remember the reader does not have the code!}

As another finding, this second \fldlib analysis also provides the complete list of locations in the source code of the
unstable branches encountered. This list has only 5 locations (which are executed multiple times due to loops), each with a unique calling context. The calling
contexts show that the first 4
locations (one of which is the unstable branch of Fig.~\ref{fig:unstable-test}) belong to the
forward wave propagation (Sect.~\ref{sec:eikonal-algo})
and that the last location belongs to the backward propagation dedicated to the path generation
(Sect.~\ref{sec:backpath-algo}).
% {\scriptsize \begin{verbatim}
% double::operator unsigned int() in mlb_rectangular.cpp:92 called by mlb_rectangular::Nearest_vector,
%   NearestIndex, eks_sethian96::set_seed, main  
% double::operator unsigned int() in mlb_rectangular.cpp:93 called by mlb_rectangular::Nearest_vector,
%   NearestIndex, eks_sethian96::set_seed, main  
% double::operator< called by priorityqueue::QueueElement::operator<, std::push_heap, std::priority_queue
%   ::push, eks_fast_marching::Loop, eks_fast_marching::run, eks_sethian96::set_seed, main  
% double::operator< called by _sethian96_hopflax::SolveQuadratic2D at eks_sethian96_hopflax.cpp:33,
%   _sethian96_hopflax::update, eks_fast_marching::Loop, eks_fast_marching::run, eks_sethian96::set_seed, main  
% double::operator> called by eks_backpath2D::update_sM, eks_sethian96::get_path, main.
% \end{verbatim}}
% 
% \commentNK{Not sure the tool output is useful,  see the above comments} 

\begin{figure}
	\centering
	\includegraphics[width=10.6cm]{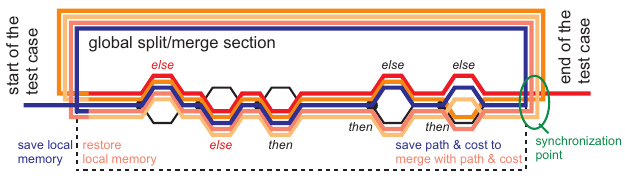}
	\caption{Trace of the \fldlib analysis forgetting unstable branches related to unsigned int conversion}
	\label{fig:fldlib-analysis-target}
\end{figure}

To investigate the impact of other unstable branches, we modify the analysis with an analysis library
that only targets the last unstable branch: the one that has a direct impact 
on the outcome of the path. The reason is that the first unstable branches listed potentially
concern all the cells of the grid of Fig.~\ref{fig:propagation} (as they are part of the forward wave propagation);
therefore, they have little chance of having an impact on the final path, whereas the last unstable
branch directly concerns the cells crossed by the resulting path
(as it belongs to the path generation step).
Concretely, the new analysis forces the ideal computation of the
unstable branch created by the \texttt{unsigned int} conversion of Fig.~\ref{fig:unstable-test}
to be equal to the floating-point point computation. Indeed, we observe in Fig.~\ref{fig:fldlib-analysis-target}
that for the first branches of the analysis paths, the control 
flow of ideal computations follows
the floating-point control flow, which was not the case in Fig.~\ref{fig:fldlib-analysis}.
The duration of the new analysis increases significantly: 17\,h\,20\,min instead of 114\,min to cover four different
branches\footnote{Here is an explanation for this time difference. The first analysis very early
encounters an unstable branch that separates the floating-point control flow from the ideal control flow.
The analysis of the floating-point control flow then propagates only constant values and the analysis
of the ideal flow stops propagating affine forms related to the difference between the float and
the ideal value since the floats are no longer present. Conversely, the new analysis has to propagate constants for
floating-point values and affine forms for ideal values and for errors during longer execution fragments, which is costly,
including the reduced product between the inferred error and the subtraction of ideal value and
floating-point value.}.

The first iteration of this third \fldlib analysis follows a joined control flow for both semantics until the
path and cost are computed, which was
not the case in the previous analysis. The reported error for this iteration on cost
is $4.71\times 10^{-13}$, which is consistent with the \cadna results ($1.323\times 10^{-15}$)
since \fldlib provides a guaranteed over-approximation while \cadna returns a
stochastic estimation of the error.
Then, as expected, the first unstable branch is found during the backward path generation.
% at the corresponding location in the source code.
% \commentNK{The reader does not have the code} 
The second analysis iteration follows only the
floating-point control flow and it gives the same result for cost as the
reference execution.
The third analysis iteration follows only the ideal control flow and the MERGE macro at the end of the
\texttt{main} function creates an error of $4.66\times 10^{-13}$ as the maximum difference between
the ideal cost and the
floating-point cost. 
An important finding here is that this small error satisfies a
sufficient stability criterion for our optimization
algorithm, even if there are only two unstable branches evaluated.

The robustness of this conservative analysis, if it is confirmed on all paths despite the unstable
branchings encountered by \cadna
and qualified by \verrou, suggests that certain computations are redone in different parts of the algorithm,
notably between the forward wave propagation (Sect.~\ref{sec:eikonal-algo}) and the backward propagation for path generation (Sect.~\ref{sec:backpath-algo}).
This would be another interesting finding for such kinds of algorithms, where some small steps are recomputed several times.
Indeed, the stochastic analysis does not ensure the introduction of exactly the same perturbation if exactly the same computation is performed several times. 
Suppose such a redundant computation evaluates to a value \textit{res} the first time, the stochastic analysis
evaluates it to $\textit{res}+\delta$ the second time, since the pertubations introduced by the
analysis are not the same.
Hence, the branch taken after the first computation may be different
from the one taken after the second, whereas the deterministic IEEE-754 computation guarantees that it
will be the same branch. This issue also occurs for \fldlib analysis in case of over-approximations.
In this case, the evaluation result is $\textit{res} + [a, b]$, but the analysis cannot guarantee that
it is the same branch because the value chosen in $[a, b]$ the first time may be different
from the value chosen in $[a, b]$ the second time. Since the intervals for the ideal values
are very very small ($319$ bits of mantissa is equivalent to a precision of $4.68\times 10^{-97}$)
and since the main linear relationships are preserved between the variables, the analysis 
is likely to avoid certain over-approximations that would consider unreachable branches and
generate false negatives.

Figure~\ref{fig:analysis-summary} shows a summary of our first experiments, which required
little investment in annotations of the source code, but a lot of effort in the definition and configuration of the analyses.
It gives an indicative (and subjective, based on our experience) 
level of confidence for the results of each tool.
The instrumentation time corresponds to the time that was required for the authors to
instrument the code. The analysis time shows the tool execution without the compilation steps.
The cost error is the error output directly produced by the tools
for the \texttt{cost} variable. It concerns the standard deviation of the cost values for
the stochastic tools (\cadna and \verrou) and the conservative error for the formally guaranteed tool (\fldlib).
The error of mean cost value is the difference between the mean of the cost values computed by the tools and
the original floating-point evaluation of the variable \texttt{cost} computed by the code without any instrumentation.
The indicative confidence in the corresponding analysis increases when both errors become closer.
The reason of error inconsistency is given in the last line.

\begin{figure}
\begin{center}
\begin{tabular}{|c|c|c|c|}
  \hline
    & \cadna & \verrou & \fldlib: $4$ paths\\
    &       &        & over $2^{351\,296}$ \\
  \hline
    instrumentation time & 10\,min & 0\,s & 3\,h \\
  \hline
    analysis time & 0.556\,s & 16.123\,s & 17\,h\,20\,min \\
  \hline
    cost error   & 1.323e-15 & 2.68e-4 & 4.71e-13 \\
  \hline
    \parbox{4cm}{\noindent\begin{center}error of mean cost value  \\ wrt. reference execution\end{center}} & 1.25e-3  & 1.27e-3 & 4.71e-13 \\
  \hline
    \parbox{4cm}{\noindent\begin{center}indicative confidence\\ in results (/10)\end{center}} & 4 & 6 & 7 \\
  \hline
    reason of error & \small unstable branching & \small original float        & \small no inconsistency \\
    inconsistency   & \small not evaluated      & \small execution not reached & \small for 4 paths \\
  \hline
\end{tabular}
\end{center}
\caption{Analysis summary (the floating-point reference execution time without analysis being 0.182\,s)}
\label{fig:analysis-summary}
\end{figure}

\subsection{Ongoing Work on Formal Verification for Thin Numerical Scenarios}

The aforementioned results are promising and make us believe that a complete formal robustness analysis for this case study is possible.
But we need to cover all of the $351\,296$ unstable branches identified by the \fldlib analysis to know
if the accuracy of the cost is rather close to 
%\texttt{1.27e-3} 
$1.27\times 10^{-3}$ 
or 
%\texttt{4.71e-13} 
$4.71\times 10^{-13}$ 
for this test case (cf. Fig.~\ref{fig:analysis-summary}). This section presents our ongoing work in this direction. 

\begin{figure}
  \centering
  \includegraphics[width=10.6cm]{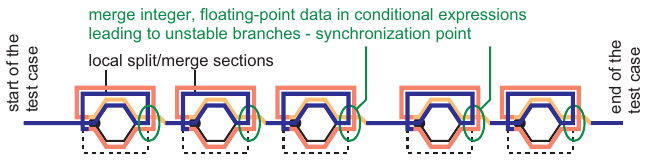}
  \caption{(Simplified) trace of the \fldlib analysis with local synchronization points}
% \commentNK{'depending from'$\to$'leading to'? or 'with'? if I understand correctly: branches depend on expressions, not the other way round, right?}
  \label{fig:fldlib-analysis-local}
\end{figure}

For this purpose, we add local synchronization annotations around the detected unstable branches (see the resulting trace in Fig.~\ref{fig:fldlib-analysis-local}). That means that
the \texttt{unsigned int} variable receiving the conversion of the floating-point computation 
in Fig.~\ref{fig:unstable-test} is conditionally defined. For instance, the evaluation of \texttt{jm_res} with the
values \texttt{y=0.5}, \texttt{yMin=0},
\texttt{dy=0.005} can be seen as producing an integer defined as $\textit{if } b_0 \textit{ then } 100 \textit{ else } 99$,
where $b_0$ is a fresh and free logical variable in $\{\textrm{true}, \textrm{false}\}$. For this unstable branch, $b_0$ evaluates to {true}
in the floating-point semantics and {false} in the ideal semantics. Further unstable
branches have a more complex evaluation in ideal semantics.

For this propagation, we define a new conditional domain\footnote{This domain is not yet available in the public repository of \fldlib, but we plan to integrate it into the open-source repository of the tool
in the near future.} in \fldlib that contains cascading
conditional expressions or a simple integer value. This domain is implemented as an instantiation of \texttt{EnhancedInteger}
template mentioned in Sect.~\ref{sec:instrumentation}. Therefore, it can represent domains 
like 
$$\textit{if } b_0 \textit{ then } (\textit{if } b_1 \textit{ then } \ldots
\textit{ else } \ldots) \textit{ else } ((\textit{if } b_2 \textit{ then } \ldots
\textit{ else } \ldots))$$

\noindent
The conditional domain also propagates to floating-point computations.
%, which suggests the creation of 
%a similar domain for floating-point variables.

The initial code contains integer and floating-point values. Our automatic instrumentation
(see Sect.~\ref{sec:instrumentation}) preserves floating-point constants but replaces
floating-point variables with the default floating-point domain containing an affine form for the ideal computation and the accumulation
of round-off errors, and an interval for the floating-point computations.
Thus, each new domain potentially interacts with 3 different domains (conditional integer,
conditional floating-point, affine forms) and the concepts of C++-20 are very useful for
handling these interactions~--- adding the conditional domains
to \fldlib required 12\,kloc of C++ code.

A finalization of these new domains and their application  for the robustness analysis of the case study is still ongoing. 
It will require a manual instrumentation of the code
(that will probably take  more than 8 hours) but 
can be expected to help analyze the target code.

\medskip

Robustness analysis is a mandatory requirement before attempting to verify functional properties,
such as the relative independence of the results with respect to the size of the grid (cf. RQ3).
Checking these properties follows the same methodology as checking robustness. We proceed first with
simple tests, then with formal analysis.
We start by using the same test case, before attempting later a modular verification approach
based on deductive methods.
%, which can also be carried out with instrumentation techniques.

%Verification by testing consists of comparing the asynchronous result of two similar executions for
%different grid sizes and checking some expected correlations between the result and the
%size.
%
%Formally verifying such a correlation in a synchronized analysis around a test case requires
%instrumentation with the template classes IntegerBranchOption and FloatingBranchOption.
%The boolean variables involved in the decision tree are the boolean variables from the robustness analysis augmented
%as weel as additional variables referring to the choice of grid size. The decision tree domain
%remains unchanged, but this requires additional instrumentation work since more types are involved
%at all stages of the analysis. This synchronized version must contain the asynchronous results
%obtained by testing.

\section{Conclusion}
\label{sec:conclusion}
% !TeX root = ./main.tex

Numerical analysis of software is important for critical programs, in particular related to trajectory computation used in autonomous systems. It is also a very 
challenging and time-consuming task. Indeed, precision and robustness of the algorithm can be impacted by many instructions, especially for
programs with long execution paths and/or simulating a continuous 
real space by a discrete grid, for which a small perturbation of data
can naturally lead to another behavior. 
 
This case study paper describes an industrial application 
of several modern numerical analysis tools
to a real-life path computation algorithm with a realistic test case. 
We present the applied methodology and results.
An important first step of the study is to ensure code instrumentability 
and to compare various analysis results to qualify the impact of unstable branches with stochastic methods (with tools like \cadna and \verrou).
Next, we investigate unstable branches and formally ensure robustness with a formal analysis (using a tool like \fldlib).
The results we obtained seem very promising: we managed to identify
the unstable branches and the corresponding locations on the code
that constitute important attention points for numerical analysis.
Dynamic analysis tools (\cadna and \verrou) show that the relative error in the path computation is sufficiently small, and the algorithm is sufficiently
robust. This conclusion should be confirmed by a formal analysis.   
A representative subset of unstable branches coming from different parts of the algorithm has been formally shown (with \fldlib) to ensure 
expected robustness properties, while the study for other branches 
is still in progress. So far, the analysis confirmed that the 
algorithm meets the user expectations in terms of accuracy and
robustness. 

\paragraph{Future Work.}
This case study suggests numerous future work perspectives.
One perspective is to finalize the investigation of unstable branches.
We plan to use the new conditional domains that were recently 
integrated into \fldlib and will be evaluated on this case study.
Considering other realistic test cases and replaying
the analyses for them is another work direction. 
Applying the described methodology on other industrial 
use cases is another perspective. 

As a more ambitious long-term research 
objective, proving that the result does not depend on the
size of the grid (RQ3) is a much more complex problem. 
Our plan is to apply
a component-based divide-and-conquer approach on the source code. 
For each component, this requires
formal instrumentation in order to propagate logical formulas instead of abstract domains. 
The starting point is the previous instrumentation of the code with its annotations for the
synchronization of unstable branches. The methodology is inspired by the approach used in
deductive verification, by first replacing the data structures of the code with classes
representing formal properties.
C++ operator overloading will propagate these properties across components using carefully designed
verification unit scenarios.
The engineer's objective will be to design unit scenarios (such as the postcondition/output invariant of
a method/class is formally contained in the precondition/input invariant of 
the method/class that takes its results).

\paragraph{Acknowledgment.}

Part of this research (tooling improvement) was supported by the ANR InterFLOP project (grant ANR-20-CE46-0009).
Part of this work was also supported by the ANR EMASS project (grant ANR-22-CE39-0014). We thank the designers and developers of numeric analysis tools
of the InterFLOP community.

\bibliographystyle{eptcs}
\bibliography{biblio}
\end{document}